\documentclass[aps,prb,twocolumn,groupedaddress,showpacs,showkeys,amsmath]{revtex4-1}
\usepackage[dvips]{graphicx}
\usepackage{dcolumn}
\usepackage{bm}
\usepackage{color}
\begin{document}

\title{High-Pressure Hydrogen Sulfide by Diffusion Quantum Monte Carlo} 

\author{Sam Azadi}
\email{s.azadi@imperial.ac.uk}
\affiliation{Department of Materials Science, Royal School of Mines, 
Thomas Young Center, London Centre for Nanotechnology, 
Imperial College London, London  SW7 2AZ, United Kingdom}
\author{Thomas D. K\"{u}hne}
\affiliation{Dynamics of Condensed Matter, Department of Chemistry,
University of Paderborn, Warburger Str. 100, D-33098 Paderborn, Germany}
\affiliation{Paderborn Center for Parallel Computing and Institute for
 Lightweight Design with Hybrid Systems, University of 
Paderborn, Warburger Str. 100, D-33098 Paderborn, Germany}

\date{\today}

\begin{abstract}
 We use the diffusion quantum Monte Carlo to revisit the enthalpy-pressure 
 phase diagram of the various products from the different proposed 
 decompositions of H$_2$S at pressures above 150~GPa. Our results entails
 a revision of the ground-state enthalpy-pressure phase diagram. Specifically,
 we find that the C2/c HS$_2$ structure is persistent up to 440~GPa before
 undergoing a phase transition into the C2/m phase. Contrary to density
 functional theory, our calculations suggest that the C2/m phase of HS
 is more stable than the I4$_1$/amd HS structure over the whole pressure
 range from 150 to 400 GPa. Moreover, we predict that the Im-3m phase is 
 the most likely candidate for H$_3$S, which is consistent with recent
 experimental x-ray diffraction measurements.
\end{abstract}
\maketitle

\section {Introduction}
Back in 1968, Ashcroft predicted that according to the BCS theory\cite{BCS},
dense hydrogen would not only be metallic\cite{WH1935}, but more importantly
also a high-temperature superconductor\cite{Ashcroft1}. Since recently it 
has been shown that dissociation is a necessary condition for the metallization
of hydrogen\cite{JETP, Azadi4}, the necessary pressure to cause bandgap closure
has remained impractical high, so that metallic hydrogen has only been realized
at finite-temperature\cite{Nellis0, Nellis1, Nellis2, Eremets, Silvera}.
Yet, an appealing way to circumvent the high pressures required to metallize hydrogen
is to precompress it in hydrogen-rich systems\cite{Ashcroft2, JFeng},
since, in general, the electronic density of states is high and the 
electron-phonon interactions are strong\cite{TAStrobel}.
 
In fact, pressurized Hydrogen-rich materials have demonstrated to be rather
promising candidates for high-T$_c$ superconductivity
\cite{Eremets08, DYKim, TScheler, XFZhou, DYKim10, GGao, GGao08, JFeng}.
In particular, Drozdov and Eremets reported that at pressures around 200~GPa,
dense hydrogen sulphide becomes metallic and superconducting with a critical
temperature ($T_c$) of 203~K\cite{APDrozdov}, which is well above the highest
T$_c$ of 133~K and 164~K that were achieved in cuprates\cite{JGBednorz} 
at ambient\cite{ASchilling} and high pressures\cite{LGao} respectively.
Recent experimental results indicate that the superconducting state 
of H$_3$S adopts a body-centered cubic (BCC) structure\cite{Einaga}. 

However, nearly all of the recent calculations on Hydrogen-rich systems are based
on the single-particle mean-field theories such as density-functional 
theory (DFT)\cite{Duan,GGao,GGao08,IErrea,IErrea13,YLi,CJPickard07,DDuan,XJChen,MMartinez}.
Even though formally exact, the exact exchange and correlation (XC)
functional is unknown from the outset and needs to be approximated,
which affects both the relative stabilities of the different crystal 
structures. Indeed, for dense hydrogen-rich materials a significant
dependence on the particular of XC functional was established
\cite{JETP, SAzadi, MAMorales, RCClay, JCP16}.  

Therefore, in this paper we revisit the stability of the individual products
 originating from the various proposed decompositions of H$_2$S for pressures
 above 150~GPa by means of highly accurate diffusion Monte Carlo (DMC) simulations.
 Using the DMC method\cite{DMC, Matthew}, the electronic many-body Schr\"{o}dinger
 equation is solved stochastically, which have yielded very accurate total energies
 for atoms\cite{Marchi, Brown}, molecules\cite{Trail, Azadi2, JCP15},
 crystals\cite{Marchi2, Mostaani, Azadi1, JCP16} including hydrogen-rich 
materials at very high pressure\cite{Neil, Azadi3, Azadi4, McMahon}.

\section {Computational Details}
At first, all of the examined structures were determined by relaxing
 the internal parameters of each phase within DFT at fixed external pressure.
 The DFT calculations were all conducted within the pseudopotential 
and plane-wave approach using the CASTEP code\cite{castep}.
 Specifically, ultrasoft pseudopotentials were employed together
 with an energy cutoff of 1000~eV\cite{ultrasoft}. The exact XC
 functional was substituted by the generalized gradient approximation 
of Perdew-Burke-Ernzerhof (PBE).\cite{PBE} The geometry and cell 
optimizations were conducted using a dense $16\times16\times16$ ${\bf k}$-point
 mesh to sample the Brillouin zone, while the nuclear forces and components 
of the stress tensor were converged to 0.01~eV/$\AA$ and 0.01~GPa, respectively. 

The \textsc{casino} code was used to perform fixed-node DMC simulations with
 a trial wave function of the Slater-Jastrow (SJ) form\cite{casino}, 
\begin{equation}
  \Psi_{\rm SJ}({\bf R}) = \exp[J({\bf R})] \det[\psi_{n}({\bf r}_i^{\uparrow})] 
  \det[\psi_{n}({\bf r}_j^{\downarrow})],
\label{eq1}
\end{equation}
where ${\bf R}$ is a $3N$-dimensional vector containing the positions of 
all $N$ electrons, ${\bf r}_i^{\uparrow}$ the position of the $i$'th spin-up electron
, ${\bf r}_j^{\downarrow}$ the position of the $j$'th spin-down electron,
 $\exp[J({\bf R})]$ a Jastrow factor, while $\det[\psi_{n}({\bf r}_i^{\uparrow})]$
 and $\det[\psi_{n}({\bf r}_j^{\downarrow})]$ are Slater determinants made of 
spin-up and spin-down one-electron wave functions. These orbitals were obtained
 from PBE-DFT calculations performed with the CASTEP plane-wave code\cite{castep},
 in conjunction with Trail-Needs Hartree-Fock pseudopotentials\cite{TN1,TN2}.
 For the purpose to approach the complete basis set limit\cite{sam}, 
a large energy cut-off of 4000~eV have been chosen. The resulting plane-wave 
orbitals were subsequently transformed into a localized ``blip''polynomial basis\cite{blip}.

The Jastrow factor within Eq.~\ref{eq1} is a positive, symmetric, explicit 
function of interparticle distances. The employed Jastrow factor includes 
polynomial one-body electron-nucleus (1b), two-body electron-electron (2b) 
and three-body electron-electron-nucleus (3b) terms, as well as plane-wave 
expansions of the electron-electron separation known as $p$ terms\cite{pterm}.
 These $p$ terms build long-ranged correlations into the Jastrow factor and thus
 significantly improve the wave function and variational energy. We also 
investigated the effect of the inhomogenous backflow (BF) coordinate transformation
 on the VMC and DMC total energies\cite{Lopez}. Our BF transformation includes 
electron-electron and electron-proton terms and is given by
\begin{equation}
\mathbf{X}_i(\{\mathbf{r}_j\})=\mathbf{r}_i+\bm{\xi}^{(e-e)}_i(\{\mathbf{r}_j\})
+\bm{\xi}^{(e-P)}_i(\{\mathbf{r}_j\}),
\end{equation}
where $\mathbf{X}_i(\{\mathbf{r}_j\})$ is the transformed coordinate of electron $i$,
 which depends on the full configuration of the system $\{\mathbf{r}_j\}$. 
The vector functions $\bm{\xi}^{(e-e)}_i(\{\mathbf{r}_j\})$ and
 $\bm{\xi}^{(e-P)}_i(\{\mathbf{r}_j\})$ are the electron-electron and electron-proton
 backflow displacements of electron $i$. They are parameterized as
\begin{equation}
\bm{\xi}^{(e-e)}_i(\{\mathbf{r}_j\})=\sum_{j\neq i}^{N_{e}} \alpha_{ij}(r_{ij}) \mathbf{r}_{ij}
\end{equation}
and
\begin{equation}
\bm{\xi}^{(e-P)}_i(\{\mathbf{r}_j\})=\sum_{I}^{N_{P}} \beta_{iI}(r_{iI})
\mathbf{r}_{iI} ,
\end{equation}
where $\alpha_{ij}(r_{ij})$ and $\beta_{iI}(r_{iI})$ are polynomial functions of
 electron-electron and electron-proton distances that contains variational parameters.
 All adjustable parameters in the Jastrow factor and backflow terms were optimized by
 minimizing the variance as well as the variational energy at the VMC level\cite{varmin1,varmin2}.
 If not explictly stated otherwise, all of our calculations were conducted using 
the SJ trail wave function including 1b, 2b, 3b and p terms augmented by the BF coordinate transformation. 

Beside the usage of twist-averaged boundary conditions (TABC) to correct 
finite-size errors\cite{TABC}, we extrapolated the energy per atom to the thermodynamic
 limit by fitting our twist-averaged DMC results for different system sizes
 to $E(N)=aN^{-b}+E(\infty)$, where $a$ and $b$ are fitting parameters and
 $E(\infty)$ is the eventual energy per atom in the infinite-system limit.
 Depending on simulation cell size, we used 8, 12 and 16 randomly chosen twists\cite{twist}. 
The enthalpy was evaluated by differentiating the polynomial fit of our the
 finite-size-corrected DMC energies as a function of volume. 

\section {Results and discussion}
\subsection{DMC Total Energies}

\begin{table}[h]
\begin{tabular}{ c c c c c }
\hline\hline 
P(GPa)&  E($N_1$)   & E($N_2$) & E($N_3$) & E($\infty$)\\
\hline 
150  & -187.6003(8) & -187.3893(6) & -187.2745(6) & -187.169(2)\\
200  & -186.9612(8) & -186.7573(8) & -186.6484(5) & -186.546(2)\\
250  & -186.3569(9) & -186.1696(8) & -186.0580(6) & -185.964(2)\\
300  & -185.7804(8) & -185.6074(6) & -185.4851(7) & -185.399(2)\\
\hline\hline 
\end{tabular}
\caption{\label{HS2C2m} Total energies of the C2/m phase of HS$_2$ at the DMC
 level of theory for four different pressures. The energies are given in 
eV/atom and are calculated for $N_1 = 48$, $N_2 = 96$ and $N_3 = 192$ particles
 in the unit cell, respectively. The extrapolated DMC energy at the infinite
 system size limit is denoted by E($\infty$).}
\end{table}
In the following, we are revisiting the crystal structures of Ref.~\onlinecite{IErrea}.
 Specifically, we begin with investigating the monoclinic C2/c and C2/m structures of HS$_2$. 
The resulting total energies as a function of pressure at the DMC level of theory for 
the two HS$_2$ structures at different system sizes and the extrapolation to the 
thermodynamic limit are shown in Table~\ref{HS2C2m} and \ref{HS2C2c}, respectively. 
\begin{table}[h]
\begin{tabular}{ c c c c c }
\hline\hline 
P(GPa)&  E($N_1$)   & E($N_2$) & E($N_3$) & E($\infty$)\\
\hline 
150  & -187.757(1) & -187.5549(8) & -187.4934(6) & -187.469(2)\\
200  & -187.254(1) & -186.9123(8) & -186.8828(5) & -186.817(2)\\  
250  & -186.767(1) & -186.3066(7) & -186.2809(6) & -186.187(2)\\ 
300  & -186.348(1) & -185.7316(7) & -185.6965(5) & -185.572(2)\\
\hline\hline 
\end{tabular}
\caption{\label{HS2C2c} The DMC total energies of the C2/c HS$_2$ structure at 
four distinct pressures. The energies are calculated for $N_1 = 24$, $N_2 = 96$
 and $N_3 = 192$ particles in the unit cell, respectively. The extrapolated DMC 
energy at the infinite system size limit is denoted by E($\infty$). All energies are in eV/atom.}
\end{table}

Comparing the DMC results for the C2/m and C2/c structures of HS$_2$, 
we find that at the same pressure, the total energy of the C2/c phase 
is throughout lower than that of the C2/m structure. Starting from 150~GPa,
 the difference is as large as 299.7~meV/atom, but is strictly decreasing to 173~meV/atom for 300~GPa. 

\begin{table}[h]
\begin{tabular}{ c c c c c }
\hline\hline 
P(GPa)&  E($N_1$)   & E($N_2$) & E($N_3$) & E($\infty$) \\
\hline 
150  & -144.6139(8) & -144.5199(5) & -144.4455(4) & -144.398(1)\\
200  & -143.9814(7) & -143.8912(6) & -143.8268(5) & -143.781(1)\\
250  & -143.2503(8) & -143.1232(6) & -143.0780(5) & -143.014(1)\\ 
300  & -142.8490(7) & -142.6977(6) & -142.6069(5) & -142.531(1)\\
\hline\hline 
\end{tabular}
\caption{\label{HSC2m} Total energies of the C2/m phase of HS at the DMC 
level of theory for four different pressures. The energies are given in 
eV/atom and are calculated for $N_1 = 64$, $N_2 = 128$ and $N_3 = 256$ 
particles in the unit cell, respectively. The extrapolated DMC energy at
 the infinite system size limit is denoted by E($\infty$).}
\end{table} 
The DMC total energies for the C2/m and I4$_1$/amd structures of HS are
 shown in Tables~\ref{HSC2m} and \ref{HSI4}, respectively. As before, all 
energies are calculated for different number of particles in the unit cell 
and extrapolated to the thermodynamic limit. 
Even though the difference in varying, the C2/m structure of HS is
 energetically throughout lower by about 500~meV/atom than the 
corresponding I4$_1$/amd phase. More precisely, at a pressure of 
150, 200, 250 and 300~GPa, the differences between the two structures
 are 580, 634, 479 and 582~meV/atom, respectively.  
\begin{table}[h]
\begin{tabular}{ c c c c c }
\hline\hline 
P(GPa)&  E($N_1$)   & E($N_2$) & E($N_3$) & E($\infty$)\\
\hline 
150  & -145.251(1) & -144.5755(8) & -144.1563(5) & -143.819(2)\\
200  & -144.988(1) & -144.1049(9) & -143.5889(5) & -143.147(2)\\
250  & -144.671(1) & -143.6350(9) & -143.0537(6) & -142.535(2)\\
300  & -144.332(1) & -143.1536(9) & -142.5379(5) & -141.948(2)\\
\hline\hline 
\end{tabular}
\caption{\label{HSI4} The DMC total energies of the I4$_1$/amd HS 
structure at four distinct pressures. The energies are calculated
 for $N_1 = 32$, $N_2 = 64$ and $N_3 = 128$ particles in the unit cell,
 respectively. The extrapolated DMC energy at the infinite system size
 limit is denoted by E($\infty$). All energies are in eV/atom.}
\end{table}

Among the various potential products of the decomposition of H$_2$S, H$_3$S 
is a particular intriguing candidate for  conventional, but high-temperature 
BCS superconductivity due to its high-frequency phonon modes. 
Interestingly, recent theoretical crystal structure prediction simulations
 suggested that at high-pressure, H$_3$S in its trigonal R3m and cubic
 Im-3m structures are the most likely products of the decomposition of H$_2$S\cite{Duan}.
 Even though chemical somewhat counter-intuitive, we also revisiting here the proposed structures by means of DMC. 

\begin{table}[h]
\begin{tabular}{ c c c c c }
\hline\hline 
P(GPa)&  E($N_1$)   & E($N_2$) & E($N_3$) & E($\infty$)\\
\hline 
150  &-80.0626(6)&-79.8131(5)&-79.7489(3)&-79.624(1)  \\
200  &-79.7696(6)&-79.5045(5)&-79.4268(3)&-79.294(1)  \\
250  &-79.4855(7)&-79.2093(4)&-79.1287(3)&-78.991(1)  \\
300  &-79.1996(7)&-78.9137(5)&-78.8325(3)&-78.689(1)  \\
\hline\hline 
\end{tabular}
\caption{\label{H3SR} Total energies of the R3m phase of H$_3$S 
at the DMC level of theory for four different pressures. The energies
 are given in eV/atom and are calculated for $N_1 = 48$, $N_2 = 96$ and
 $N_3 = 192$ particles in the unit cell, respectively. The extrapolated 
DMC energy at the infinite system size limit is denoted by E($\infty$).}
\end{table}
The corresponding total energies, as computed by DMC, are listed in Tables~\ref{H3SR}
 and ~\ref{H3SI}, respectively. 
\begin{table}[h]
\begin{tabular}{ c c c c c }
\hline\hline 
P(GPa)&  E($N_1$)   & E($N_2$) & E($N_3$) & E($\infty$)\\
\hline 
150  &-79.3944(5)&-79.6212(4)&-79.6717(3)&-79.785(1) \\
200  &-79.0955(5)&-79.3326(4)&-79.3767(3)&-79.495(1)  \\
250  &-78.7941(5)&-79.0420(4)&-79.0851(3)&-79.209(1)  \\
300  &-78.4924(4)&-78.7555(4)&-78.7934(3)&-78.925(1)  \\
\hline\hline 
\end{tabular}
\caption{\label{H3SI} The DMC total energies of the Im-3m pH$_3$S 
structure at four distinct pressures. The energies are calculated
 for $N_1 = 64$, $N_2 = 128$ and $N_3 = 256$ particles in the unit cell,
 respectively. The extrapolated DMC energy at the infinite system size 
limit is denoted by E($\infty$). All energies are in eV/atom.}
\end{table}

We find that in thermodynamic limit, the Im-3m structure is
 energetically more favorable than the R3m phase of H$_3$S over
 the whole pressure range considered here, and that the difference
 slightly widens with increasing pressure. 
 
\subsection{Backflow Wavefunction}
However, due to the necessary fixed-node approximation in order to cope 
with the infamous fermion sign-problem\cite{FSWF1, FSWF2}, the fixed-node 
DMC method samples the variationally optimal many-electron wave function,
 which is consistent with an \textit{a priori} given nodal surface of a 
presumed trial wave function, instead of the exact ground-state wave function\cite{DMC}.
 [The nodal surface of an $N$-electron wave function $\Psi(\bm{r}_1,\bm{r}_2,\ldots,\bm{r}_N)$
 is the $(3N-1)$-dimensional hypersurface on which $\Psi$ is zero.] Therefore, 
the accuracy of the presumed trial wave function, determines the quality the 
eventual results via the nodal surface, which represents the sole approximation of
 the employed fixed-node DMC method.

As already alluded to previously, using the so called BF coordinate 
transformation\cite{BF1, BF2, BF3}, the orbitals in the Slater determinant
 are evaluated not at the actual electron positions, but on quasi-electron
 positions that are functions of all the particle coordinates. However, 
the BF function, which describes the offset of the quasi-electron coordinates
 relative to the actual coordinates, contains free parameters, which are
 determined by a variational optimization of the trial wave function. 
In this way, the nodes of the BF trial wave function are no longer fixed,
 but do have some flexibility to  move during the trial wave function 
optimization in order to further minimize the variational energy.

\begin{figure}
\centering
\includegraphics[width=0.5\textwidth]{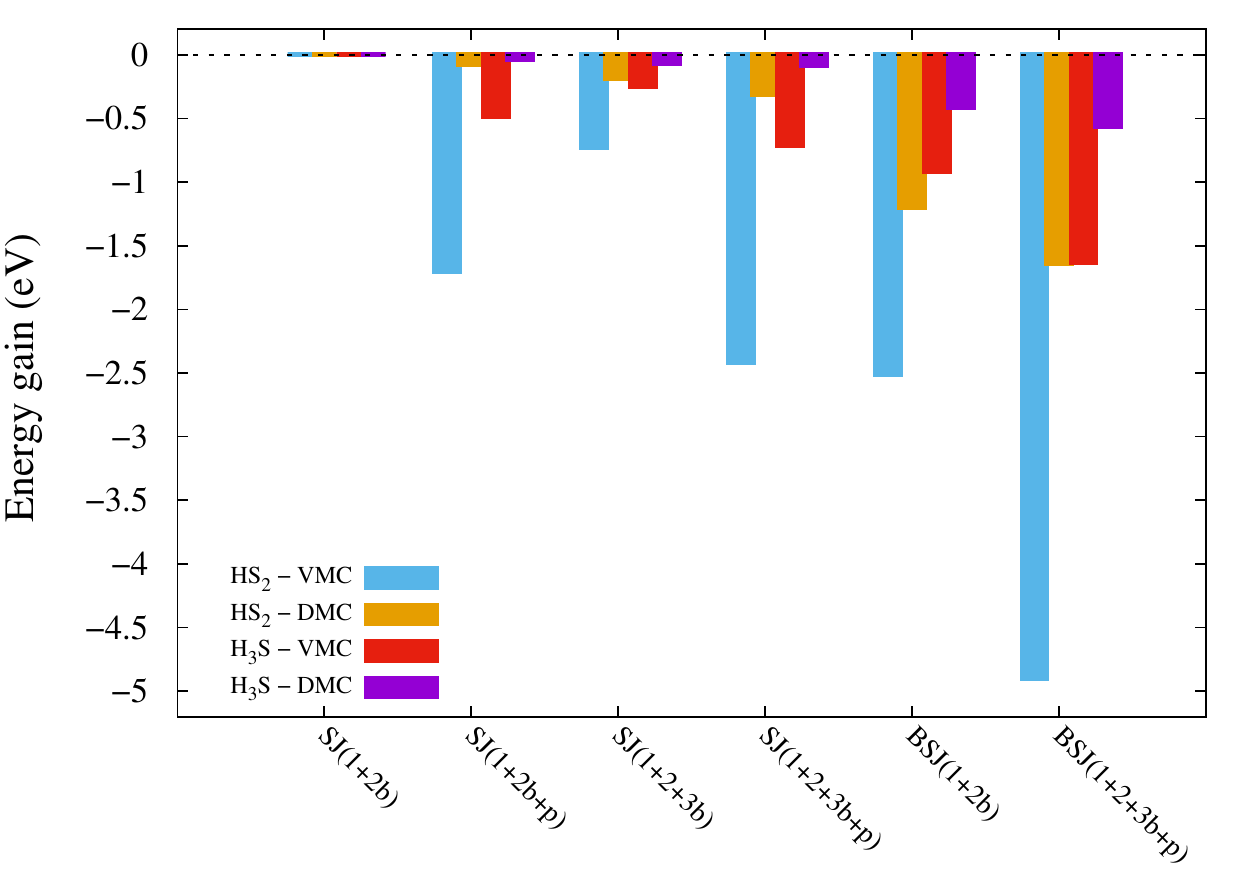}
\caption{\label{WFs}(colour online) The reduction of variational 
energy for the HS$_2$ and H$_3$S systems using different trial wave 
functions with respect to the SJ(1b+2b) approach using the VMC and DMC methods.} 
\end{figure}

\begin{table}[ht]
\begin{tabular}{ c c c c }
\hline\hline
WF               & VMC        &  DMC       & variance \\ \hline
SJ(1b+2b)        & -81.9882(5)& -82.1819(5)& 6.97(2)   \\
SJ(1b+2b+$p$)    & -82.0508(3)& -82.1848(5)& 6.22(4)  \\
SJ(1b+2b+3b)     & -82.0150(4)& -82.1886(5)& 6.54(3)  \\
SJ(1b+2b+3b+$p$) & -82.0772(3)& -82.1932(4)& 5.82(2)  \\
BSJ(1b+2b)       & -82.0806(3)& -82.2259(4)& 5.10(3)  \\
BSJ(1b+2b+3b+$p$)& -82.1683(3)& -82.2423(4)& 4.17(4)  \\
\hline\hline
\end{tabular}
\caption{\label{HS2_WF} The VMC and DMC total energies for the C2/m phase
 of HS$_2$ as calculated using the SJ(1b+2b), SJ(1b+2b+$p$),
 SJ(1b+2b+3b), SJ(1b+2b+3b+$p$), as well as BSJ(1b+2b) and BSJ(1b+2b+3b+$p$)
 trial wave functions. All energies are in Hartree per primitive unit cell.} 
\end{table}
Here, we have employed a large variety of different trial wave functions
 within our VMC and DMC energies of HS$_2$ and H$_3$S, which are the systems
 with of lowest and highest hydrogen densities we have considered. 
Specifically, we applied SJ-type trial wave functions including one- and two-body
 terms (SJ(1b+2b)), an additional three-body term (SJ(1b+2b+3b)), 
as well as the respective versions that are augmented by a $p$-term denoted 
as SJ(1b+2b+$p$) and SJ(1b+2b+3b+$p$), respectively. In addition, 
Backflow-Slater-Jastrow (BSJ) trial wave functions including one- and two-body
 terms (BSJ(1b+2b)) and the subsequent variant including an additional three-body
 and the $p$ term (BSJ(1b+2b+3b+$p$)) are considered. The resulting VMC and DMC 
total energies for the C2/m phase of HS$_2$ and Im-3m H$_3$S structure are listed 
in Tables~\ref{HS2_WF} and \ref{H3S_WF}, respectively. 
\begin{table}[ht]
\begin{tabular}{ c c c c }
\hline\hline
WF               & VMC        &  DMC       & variance \\ \hline
SJ(1b+2b)        & -23.1044(3)& -23.1650(2)& 4.21(4)  \\
SJ(1b+2b+$p$)    & -23.1223(2)& -23.1663(2)& 3.75(1)  \\
SJ(1b+2b+3b)     & -23.1134(2)& -23.1674(3)& 3.90(2)  \\
SJ(1b+2b+3b+$p$) & -23.1307(2)& -23.1680(2)& 3.54(1)  \\
BSJ(1b+2b)       & -23.1381(2)& -23.1803(3)& 2.81(2)  \\
BSJ(1b+2b+3b+$p$)& -23.1645(2)& -23.1856(1)& 2.278(9) \\
\hline\hline
\end{tabular}
\caption{\label{H3S_WF} The VMC and DMC total energies of the Im-3m
 H$_3$S structure as calculated using the SJ(1b+2b), SJ(1b+2b+$p$),
 SJ(1b+2b+3b), SJ(1b+2b+3b+$p$), as well as BSJ(1b+2b) and BSJ(1b+2b+3b+$p$)
 trial wave functions. All energies are in Hartree per primitive unit cell.} 
\end{table}
 The energy gain of the various trial wave functions with respect to 
 the SJ(1+2b) approach using the VMD and DMC methods for the two different
 systems are shown in Fig.~\ref{WFs}. It is apparent that the addition 
 of the three-body term and more so the $p$ term substantially reduces
 the VMC energy, but tht the DMC is only marginally affected since the
 initial nodal structure is identical. The usage of the BF transformation,
 however, improves the nodal surface and hence entails an energy 
 lowering at the VMC and the DMC levels. Interestingly, the gain in energy
 for HS$_2$ is much more pronounced than for H$_3$S, which demonstrates 
 the importance of an accurate trail wave function for the former. 
 For the Im-3m phase of H$_3$S the energy gain is -1.634 and -0.56~eV 
 at the VMC and DMC levels of theory, while for the C2/m HS$_2$ structure, 
 the energy gain can be as high as -4.898 and -1.643~eV. 
 The increased accuracy of the BF wave function, however, comes at a
 rather high computational cost, which is due to the necessity to
 evaluate the orbitals and their first two derivatives and moreover 
 also the collective BF coordinates, because every element of the Slater
 matrix must be updated every time a single electron is moved. 
 Even though it is partially compensated by the fact that the less 
 complex BSJ(1b+2b) trail wave function is equally accurate than the
 much more complex SJ(1b+2b+3b+$p$) calculations, we have elected to
 use the former instead of the latter in the following DMC calculations 
 of the enthalpy-pressure phase diagram calculations. 
\subsection{DMC Enthalpy-Pressure Phase Diagram}
 In order to compute the enthalpy-pressure phase diagrams for the 
 different structures, we fitted our extrapolated DMC total energies 
 as a function of volume $V$ against a model equations of state $E(V)$. 
 We found that a quadratic polynomial is a sufficiently accurate representation
 of our actual DMC energies. Using this model, it is straightforward to calculate
 the pressure $P(V)=-dE(V)/dV$ as a function of $V$ and thus also the enthalpy
 per atom $H = E + PV$. 

\begin{figure}
\centering
\includegraphics[width=0.4\textwidth]{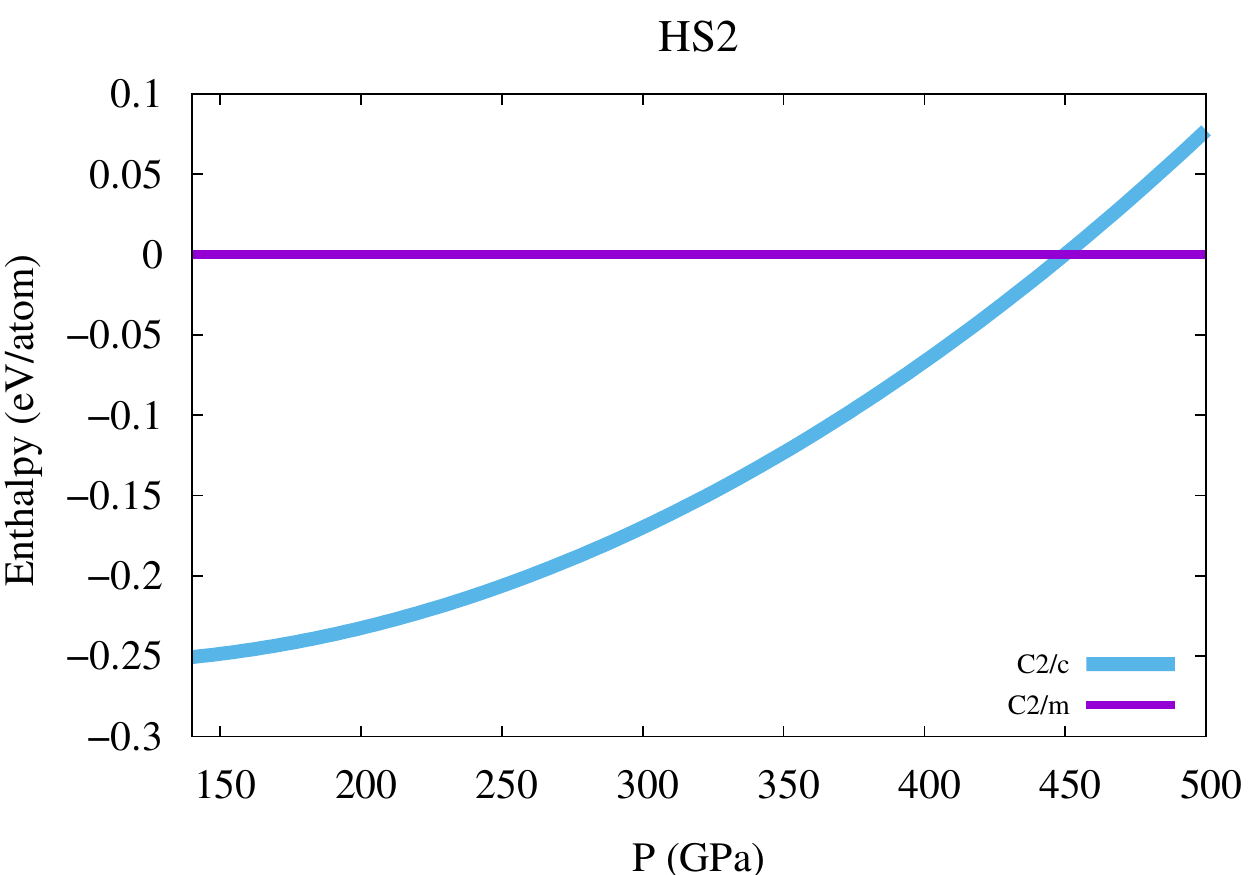}
\caption{\label{HPHS2}(colour online) The DMC enthalpy as a function 
of pressure of the C2/c HS$_2$ structure relative to the C2/c phase. 
The estimated uncertainties in the DMC enthalpies due to statistical 
and systematic errors are represented by the widths of the corresponding lines.} 
\end{figure}

 In previous DFT calculations including ZPE correction it was predicted 
 that at 200~GPa 5H$_2$S decomposes into 3H$_3$S and HS$_2$, where HS$_2$
 adopts C2/c symmetry, but undergoes a phase transition to the more stable
 C2/m structure at 250~GPa\cite{IErrea}. However, the present DMC enthalpies
 indicate that the C2/c HS$_2$ structure is more stable than the C2/m 
 phase up to 440~GPa, as it is shown in Fig.~\ref{HPHS2}. 
 In fact, the enthalpy difference between the C2/c and C2/m phases
 of HS$_2$ is much larger than the ZPE correction as estimated by DFT\cite{IErrea}. 

\begin{figure}
\centering
\includegraphics[width=0.4\textwidth]{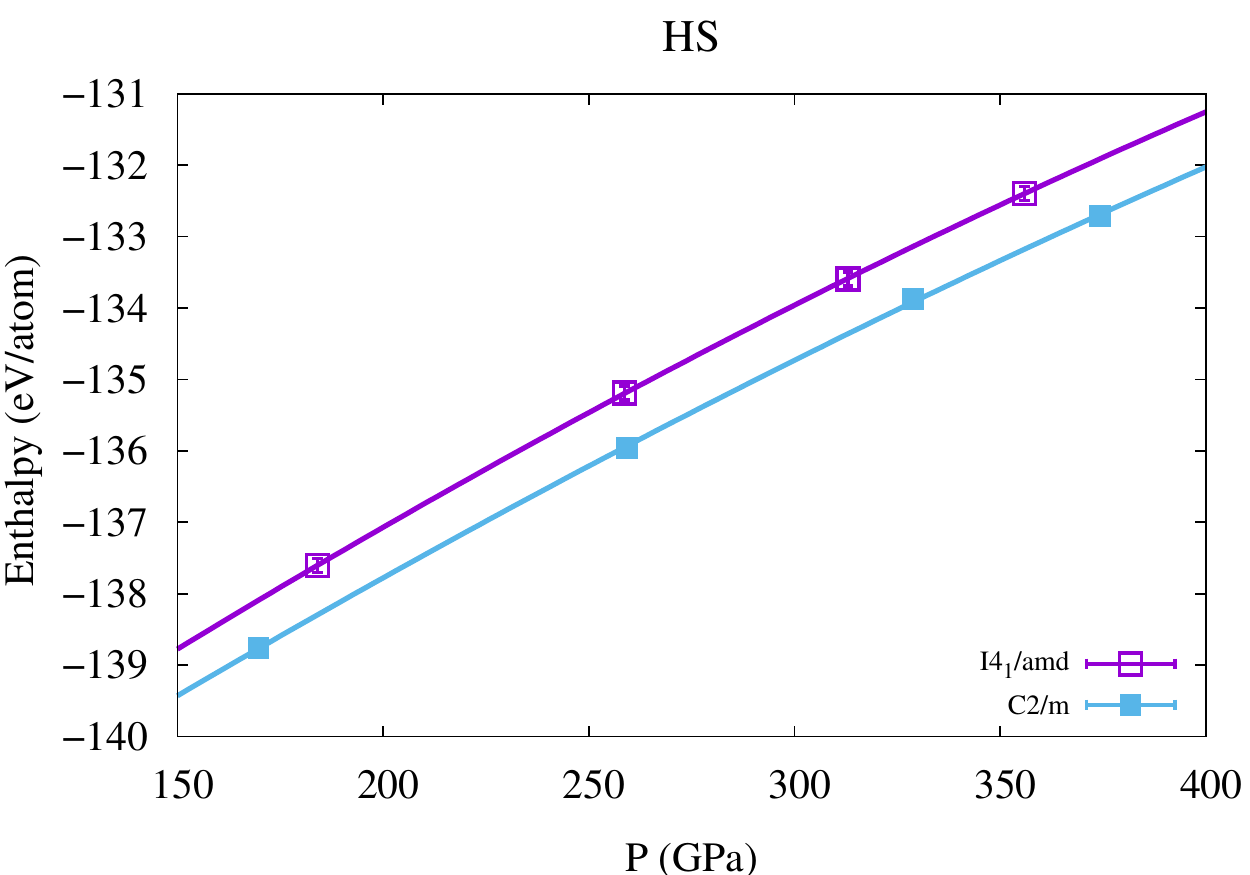}
\caption{\label{HPHS}(colour online) The DMC enthalpy of the C2/m and
 I4$_1$/amd phases of HS as a function of pressure. The corresponding 
error bars are smaller than the size of the data points.} 
\end{figure}
 Moreover, DFT-based crystal structure searches predict that the
 energetically most favorable phase of HS at 200~GPa is I4$_1$/amd
 and at 300~GPa C2/m\cite{IErrea}. The present DMC enthalpy-pressure
 curves of HS are shown in Fig.~\ref{HPHS}. At variance to DFT,
 they suggest that the C2/m phase of HS is more stable than the 
 I4$_1$/amd HS structure over the whole pressure range from 150
 to 400~GPa. In fact, the difference in enthalpy between the C2/m 
 and I4$_1$/amd phases is more than 600 meV/atom and is even larger
 with increasing pressure. 

\begin{figure}
\centering
\includegraphics[width=0.4\textwidth]{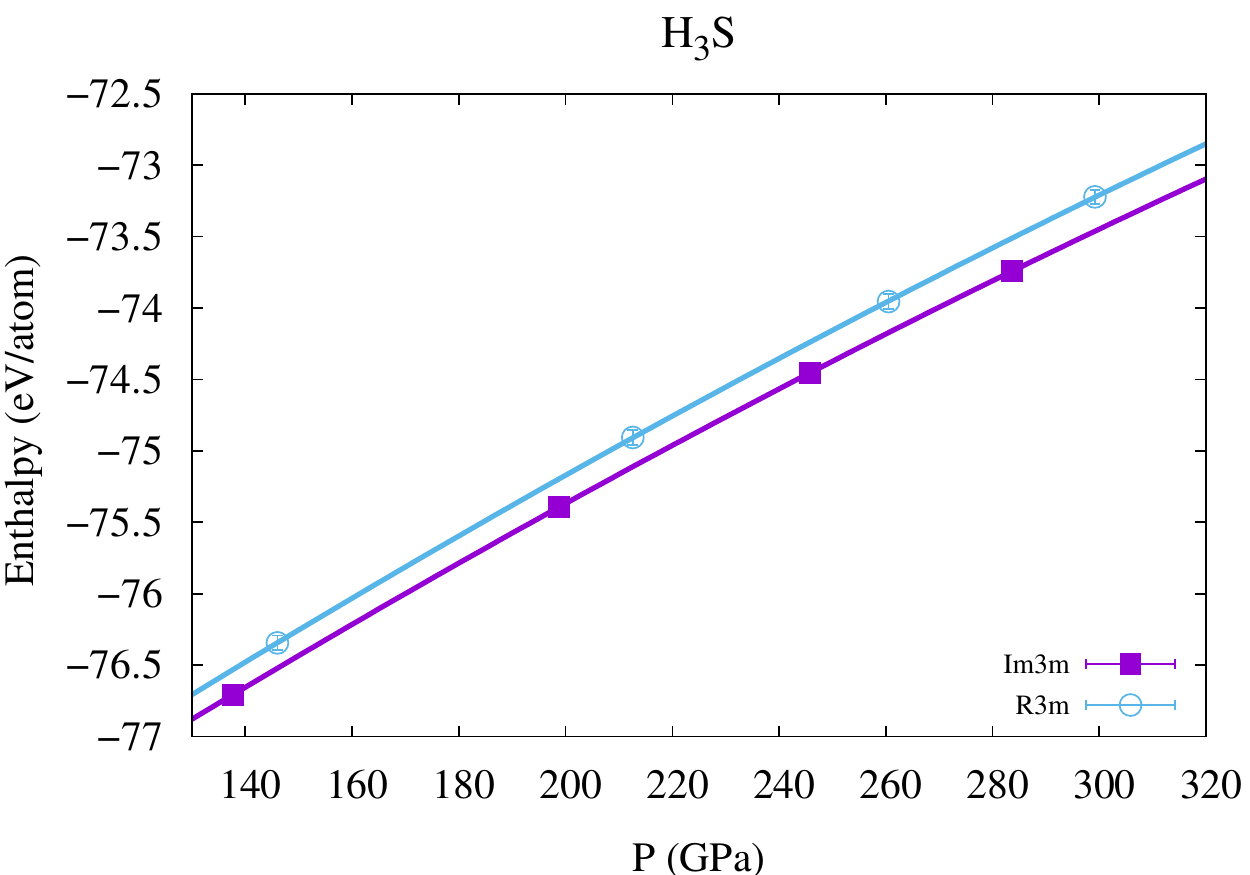}
\caption{\label{HPHS}(colour online) The DMC enthalpy of the Im-3m 
and R3m phases of H$_3$S as a function of pressure. The corresponding 
error bars are smaller than the size of the data points.} 
\end{figure}

 Previous crystal structure prediction calculations using DFT indicate
 that the R3m H$_3$S structure is stable at 130~GPa and that the Im-3m
 phase of H$_3$S is the most likely candidate at pressures larger than 200~GPa.
 Yet, our DMC enthalpy-pressure calculations again show that the Im-3m H$_3$S 
 structure is the more favorable candidate up to at least 320~GPa, which is in
 agreement with very recent experimental results.\cite{Einaga} 

 This is to say that altogether, our DMC results leads to a revision of the
 enthalpy-pressure phase diagram between 150 and 320~GPa. Specifically, 
 we find that the C2/c HS$_2$, C2/m HS and Im-3m H$_3$S structures are
 energetically most favorable within the considered pressure range. 
 However, it is well-known that the ZPE corrections plays an important 
 role in the numerical determination of the phase diagram hydrogen-rich 
 systems\cite{SAzadi,Azadi3,Azadi4}. Nevertheless, DFT calculations of 
 others have shown that the energy difference between H$_2$S and S+H$_2$
 due to ZPE is about 6 meV/f.u. at 160~GPa\cite{YLi}, which is much 
 smaller than our DMC enthalpy differences. We thus predict that the
 effect of ZPE on our DMC results is negligible. 

\section{Conclusion}
In summary, using highly accurate DMC calculations, we find 
that in the pressure range between 150 and 320~GPa, 
the C2/c HS$_2$, C2/m HS and the superconducting Im-3m H$_3$S
 structures are the enthalpically most favorable products of 
the decomposition of H$_2$S. 
Nevertheless, we conclude by noting that instead of the often
 proposed decomposition of H$_2$S into H$_3$S, HS$_2$, HS, or 
rather elemental $S$, the dissociation into H$_3$S$^+$ and HS$^-$
 would be chemically much more sensible and still compatible with
 the recently performed XRD measurements\cite{Einaga}. 

\begin{acknowledgments}
This work made use of computing facilities provided by ARCHER, 
the UK national super computing service, and by the High Performance 
Computing Service of Imperial College London, as well as of the OCuLUS
 system of the Paderborn Center for Parallel Computing (PC$^2$).
 In particular, we acknowledge that the results of this research
 have been achieved using the PRACE-3IP project (FP7 RI-312763)
 resource ARCHER based in UK. Support from the Thomas Young Centre
 under grant number TYC-101 and EP/N50869X/1 is also thankfully acknowledged. 
\end{acknowledgments}

\end{document}